\documentclass[10pt,conference,a4paper]{IEEEtran}
\usepackage{times,amsmath,epsfig}
\usepackage{amsfonts}
\usepackage{amsbsy}
\usepackage{amsmath,amssymb}
\usepackage{times}
\usepackage{graphicx}
\usepackage{latexsym}
\title {\huge Adaptive Minimum BER Reduced-Rank Linear Detection for Massive MIMO Systems
\vspace{-0.65em} }

\author
{%
{Yunlong Cai{\small $~^{\#1}$} and Rodrigo C. de Lamare{\small $~^{*2}$} }%
\vspace{1.6mm}\\
\fontsize{10}{10}\selectfont\itshape
$^{\#}$\,Department of Information Science and Electronic Engineering, Zhejiang University,  Hangzhou 310027, China\\
\fontsize{9}{9}\selectfont\ttfamily\upshape

\vspace{1.2mm}
\fontsize{10}{10}\selectfont\rmfamily\itshape

$^{*}$\,Department of Electronics, University of York, York, UK, YO10 5DD\\
\fontsize{9}{9}\selectfont\ttfamily\upshape
Emails: ylcai@zju.edu.cn, rcdl500@ohm.york.ac.uk \vspace{-1.2em} }

\begin{document}
\maketitle \thispagestyle{empty}

\begin{abstract}
In this paper, we propose a novel adaptive reduced-rank strategy for
very large multiuser multi-input multi-output (MIMO) systems. The
proposed reduced-rank scheme is based on the concept of joint
iterative optimization (JIO) of filters according to the
minimization of the bit error rate (BER) cost function. The proposed
optimization technique adjusts the weights of a projection matrix
and a reduced-rank filter jointly. We develop stochastic gradient
(SG) algorithms for their adaptive implementation and introduce a
novel automatic rank selection method based on the BER criterion.
Simulation results for multiuser MIMO systems show that the proposed
adaptive algorithms significantly outperform existing schemes.\\
\hspace{-8pt} \footnote{This work is  supported by the Fundamental
Research Funds for the Central Universities and the NSF of China
under Grant 61101103. }

\emph{Index Terms}- Multiuser MIMO systems, massive MIMO,
reduced-rank methods, adaptive algorithms, BER cost function.

\end{abstract}

\section{Introduction}

Wireless communication research has recently focused on multi-input
multi-output (MIMO) systems in order to exploit the increased
capacity offered by the use of multiple antennas, and improve the
quality and reliability of wireless links \cite{tse}. In MIMO
systems, two configurations can be employed, namely, diversity and
spatial multiplexing, which exploit spatial diversity to combat
fading and increase the data rates by transmitting independent data
streams, respectively. In particular, spatial multiplexing can be
used for multiuser MIMO systems to transmit multiple data streams
that can be separated using signal processing techniques at the
receiver. More recently, multiuser detection has been considered in
conjunction with MIMO techniques, which is widely believed to play
an important role in future communication systems
\cite{choi,rontogiannis,peng}. A recent trend has been introduced
with the concept of massive MIMO \cite{marzetta} and the
investigation of algorithms for very large MIMO systems
\cite{murch,chockalingam}, which present key technical challenges
for designers. Central problems in very large multiuser MIMO systems
are the tasks of detection and parameter estimation that are
required for interference suppression and must deal with a large
number of parameters.

In this context, reduced-rank signal processing is a very promising
technique due to its ability to deal with a large number of
parameters. It has received significant attention in the past
several years, since it provides faster convergence speed, better
tracking performance and an increased robustness against
interference as compared to conventional schemes operating with a
large number of parameters. A number of reduced-rank techniques have
been developed to design the subspace projection matrix and the
reduced-rank filter \cite{eign}-\cite{jidf}. Among the first schemes
are eigendecomposition-based (EIG) algorithms  \cite{eign},
\cite{eign2}. The multistage Wiener filter (MWF) has been
investigated in \cite{MWF2} and \cite{MWF}, whereas the auxiliary
vector filtering (AVF) algorithm has been considered in \cite{avf1}.
EIG, MWF and AVF have faster convergence speed with a much smaller
filter size, but their computational complexity is very high. A
strategy based on the joint and iterative optimization (JIO) of a
subspace projection matrix and a reduced-rank filter has been
reported in \cite{delamarespl07,delamaretvt10}, whereas algorithms
with switching mechanisms have been considered in \cite{jidf} for
DS-CDMA systems.

Most of the contributions to date are either based on the
minimization of the mean square error (MSE) and/or the minimum
variance criteria \cite{eign}-\cite{jidf}, which are not the most
appropriate  metric from a performance viewpoint in digital
communications. Design approaches that can minimize the bit error
rate (BER) have been reported in \cite{mber2,delamare_mber,mber3}
and are termed adaptive minimum bit error rate (MBER) techniques.
The work in \cite{mber3} appears to be the first approach to combine
a reduced-rank algorithm with the BER criterion. However, the scheme
is a hybrid  between an EIG or an MWF approach, and a BER scheme in
which only the reduced-rank filter is adjusted in an MBER fashion.

In this paper, we propose adaptive reduced-rank techniques based on
a novel JIO strategy that minimizes the BER cost function for very
large multiuser MIMO systems. The proposed strategy adjusts the
weights of both the rank-reduction matrix and the reduced-rank
filter jointly in order to minimize the BER. We develop stochastic
gradient (SG) algorithms for their adaptive implementation and
present an automatic rank selection method with the BER as a metric.
Simulation results for large multiuser MIMO systems show that the
proposed algorithms significantly outperform existing schemes.

The paper is structured as follows. Section \ref{Section2:system}
briefly describes the multiuser MIMO system model. The derivation of the MBER
reduced-rank algorithm
 is
described in section \ref{Section3:designmber}.
The complexity analysis of the proposed algorithm and the automatic rank selection scheme are introduced
in section \ref{Section4:proposedmber}.
 The
simulation results are presented in section
\ref{Section5:simulations}. Finally, section
\ref{Section6:conclusion} draws the conclusions.

\section{System Model}
\label{Section2:system}

Let us consider the uplink of an uncoded synchronous multiuser MIMO
system with $K$ users and one base station (BS), where each user is
equipped with a single antenna and the BS is with $M$ uncorrelated
receive antennas, $K\leq M$. We assume that the channel is a MIMO
time-varying flat fading channel.
%
The
$M$-dimensional received vector is given by
\begin{equation}
\begin{split}
\mathbf {r}(i) & =  \sum_{k=1}^{K}A_{k}\mathbf{h}_{k}(i)b_{k}(i) + \mathbf {n}(i),
\end{split}
\end{equation}
where  $b_{k}(i)$ $\in$ $\{\pm1\}$ is the $i$-th
symbol for user $k$, and the amplitude of user $k$ is $A_{k}$, $k=1,\ldots,K$. The
$M \times 1$ vector $\mathbf{h}_{k}(i)$ is the
channel vector of user $k$,
 which  is given by
 \begin{equation}
 \mathbf {h}_{k}(i)=[h_{k,1}(i) \ldots
h_{k,M}(i)]^{T},
 \end{equation}
 %
whose elements $h_{k,f}(i)$, $f=1,\ldots, M$, are independent and
identically distributed complex Gaussian variables with zero mean
and unit variance, $\mathbf {n}(i) = [n_{1}(i) \ldots n_{M}(i)]^T$
is the complex Gaussian noise vector with zero mean and $E[\mathbf
{n}(i)\mathbf {n}^{H}(i)] = \sigma^2 \mathbf {I}$, where $\sigma^2$
is the noise variance,  $(.)^T$ and $(.)^H$ denote transpose and
Hermitian transpose, respectively.

\section{Design of MBER Reduced-Rank Schemes}
\label{Section3:designmber}

In this section, we detail the design of reduced-rank schemes which
minimize the BER. In a reduced-rank algorithm, an $M \times D$
projection matrix $\mathbf{S}_{D}$ is applied to the received data
to extract the most important information of the processed data by
performing dimensionality reduction, where $1 \leq D \leq M$. A $D
\times 1$ projected received vector is obtained as follows
\begin{equation}
\mathbf {\bar{r}}(i)=\mathbf {S}^{H}_{D}\mathbf {r}(i),
\end{equation}
where it is the input to a $D\times 1$ filter $\mathbf
{\bar{w}}_{k}=[\bar{w}_{1}, \bar{w}_{2}, \ldots, \bar{w}_{D}]^T$. The
filter output is given by
\begin{equation}
\bar{x}_{k}(i)=\mathbf {\bar{w}}^{H}_{k}\mathbf
{\bar{r}}(i)=\mathbf {\bar{w}}^{H}_{k} \mathbf {S}^{H}_{D}\mathbf
{r}(i).
\end{equation}
%
The estimated symbol of user $k$ is given by
\begin{equation}
\hat{b}_{k}(i)=\textrm {sign}\{\Re[\mathbf{\bar{w}}^{H}_{k}\mathbf{\bar{r}}(i)]\},
\end{equation}
where the operator $\Re[.]$ retains the real part of the
argument and $\textrm {sign} \{.\}$ is the signum function.
The probability of error for user $k$ is given by
\begin{equation}
\begin{split}
P_{e} &= P(\tilde{x}_{k}<0)=\int^{0}_{-\infty}f(\tilde{x}_{k}) d\tilde{x}_{k}\\&= Q \bigg( \frac{\textrm{sign} \{ b_{k}(i)\}\Re[\bar{x}_{k}(i)]}{\rho (\mathbf{\bar{w}}^{H}_{k}\mathbf{S}_{D}^{H}\mathbf{S}_{D}\mathbf{\bar{w}}_{k})^{\frac{1}{2}}} \bigg),\label{eq:proberror}
\end{split}
\end{equation}
where $\tilde{x}_{k}=\textrm{sign} \{
b_{k}(i)\}\Re[\bar{x}_{k}(i)]$, $f(\tilde{x}_{k})$ is the single
point kernel density estimate \cite{mber2} which is given by
\begin{equation}
\begin{split}
f(\tilde{x}_{k})=&\frac{1}{\rho \sqrt{2\pi \mathbf{\bar{w}}^{H}_{k}\mathbf{S}_{D}^{H}\mathbf{S}_{D}\mathbf{\bar{w}}_{k} }}
 \\& \times \exp \bigg(
\frac{-(\tilde{x}_{k}-\textrm{sign}  \{ b_{k}(i)\}\Re[\bar{x}_{k}(i)])^2}{2\mathbf{\bar{w}}^{H}_{k}\mathbf{S}_{D}^{H}\mathbf{S}_{D}\mathbf{\bar{w}}_{k}\rho^2}\bigg),
\end{split}
\end{equation}
where $\rho$ is the radius parameter of the kernel density estimate,
$Q(.)$ is the Gaussian error function. The parameters of $\mathbf
{S}_D$ and $\mathbf{ \bar{w}}_{k}$ are designed to minimize the
probability of error. By taking the gradient of (\ref{eq:proberror})
with respect to $\mathbf {\bar{w}}^{*}_{k}$ and after further
mathematical manipulations we obtain
\begin{equation}
\begin{split}
\frac{\partial P_{e}}{\partial \mathbf{\bar{w}}^{*}_{k}} & =
\frac{-\exp \bigg(\frac{-|\Re[\bar{x}_{k}(i)]|^2}{2\rho^2 \mathbf
{\bar{w}}^{H}_{k} \mathbf {S}_{D}^{H} \mathbf {S}_{D} \mathbf
{\bar{w}}_{k}} \bigg)}{\sqrt{2\pi}} \times \frac{\partial \bigg(
\frac{\textrm{sign} \{ b_{k}(i)\}\Re[\bar{x}_{k}(i)]} {\rho
(\mathbf {\bar{w}}^{H}_{k} \mathbf {S}_{D}^{H} \mathbf {S}_{D}
\mathbf {\bar{w}}_{k})^{\frac{1}{2}} } \bigg)}{\partial \mathbf
{\bar{w}}^{*}_{k}}
\\&=\frac{-\exp \bigg(\frac{-|\Re[\bar{x}_{k}(i)]|^2}{2\rho^2 \mathbf {\bar{w}}^{H}_{k}\mathbf {S}_{D}^{H}\mathbf {S}_{D}
\mathbf {\bar{w}}_{k}} \bigg)\textrm{sign} \{ b_{k}(i)\}}{2\sqrt{2\pi}\rho}
\\&\times \bigg( \frac{\mathbf {S}^{H}_{D}\mathbf {r}}{(\mathbf {\bar{w}}^{H}_{k}
\mathbf {S}_{D}^{H} \mathbf {S}_{D} \mathbf
{\bar{w}}_{k})^{\frac{1}{2}}}-\frac{\Re[\bar{x}_{k}(i)] \mathbf
{S}^{H}_{D} \mathbf {S}_{D} \mathbf {\bar{w}}_{k}}{( \mathbf
{\bar{w}}^{H}_{k} \mathbf {S}_{D}^{H}\mathbf {S}_{D} \mathbf
{\bar{w}}_{k})^{\frac{3}{2}}}\bigg)\label{eq:proberror2}.
\end{split}
\end{equation}
By taking the gradient of (\ref{eq:proberror}) with respect to
$\mathbf {S}^{*}_{D}$ and following the same approach we have
\begin{equation}
\begin{split}
\frac{\partial P_{e}}{\partial \mathbf {S}^{*}_{D}} & =\frac{-\exp
\bigg(\frac{-|\Re[\bar{x}_{k}(i)]|^2}{2\rho^2\mathbf{
\bar{w}}^{H}_{k}\mathbf {S}_{D}^{H} \mathbf {S}_{D} \mathbf{
\bar{w}}_{k}} \bigg)}{\sqrt{2\pi}} \times \frac{\partial \bigg(
\frac{\textrm{sign} \{ b_{k}(i)\}\Re[\bar{x}_{k}(i)]}{\rho
(\mathbf{\bar{w}}^{H}_{k} \mathbf {S}_{D}^{H} \mathbf {S}_{D}
\mathbf {\bar{w}}_{k})^{\frac{1}{2}}} \bigg)} {\partial \mathbf{
S}^{*}_{D}}
\\&=\frac{-\exp \bigg(\frac{-|\Re[\bar{x}_{k}(i)]|^2}{2\rho^2 \mathbf {\bar{w}}^{H}_{k} \mathbf {S}_{D}^{H}\mathbf {S}_{D} \mathbf {\bar{w}}_{k}} \bigg)\textrm {sign}\{ b_{k}(i)\}}{2\sqrt{2\pi}\rho}
 \\&\times \bigg(\frac{\mathbf {r}\mathbf{ \bar{w}}^{H}_{k}}{(\mathbf {\bar{w}}^{H}_{k}\mathbf {S}_{D}^{H}\mathbf {S}_{D}\mathbf{ \bar{w}}_{k})^{\frac{1}{2}}}-\frac{\mathbf {S}_{D}\mathbf {\bar{w}}_{k}
\mathbf {\bar{w}}^{H}_{k}\Re[\bar{x}_{k}(i)]}{(\mathbf
{\bar{w}}^{H}_{k}\mathbf {S}_{D}^{H}\mathbf {S}_{D} \mathbf{
\bar{w}}_{k})^{\frac{3}{2}}} \bigg)\label{eq:proberror3}.
\end{split}
\end{equation}

\section{Proposed MBER Adaptive Algorithms}
\label{Section4:proposedmber}

In this section, we firstly describe the proposed scheme and MBER adaptive SG
algorithms to adjust the weights of $\mathbf {S}_D(i)$ and $\mathbf
{\bar{w}}(i)$ based on the minimization of the BER criterion. Then, a
method for automatically selecting the rank of the algorithm using
the BER criterion is presented.

\begin{figure}[!hhh]
\centering \scalebox{0.54}{\includegraphics{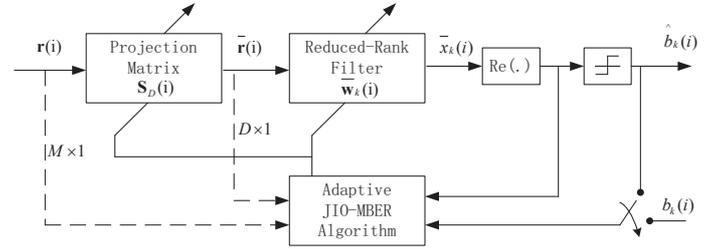}}
\vspace{-0.95em}\caption{Structure of the proposed reduced-rank
scheme } \label{fig:jiomberblock}
\end{figure}

\subsection{ Adaptive Estimation of Projection Matrix and Receiver}

The proposed scheme is depicted in Fig. \ref{fig:jiomberblock}, the
projection matrix $\mathbf {S}_{D}(i)$ and the reduced-rank filter
$\mathbf {\bar{w}}_{k}(i)$ are jointly optimized according to the
BER criterion. The algorithm has been devised to start its operation
in the training (TR) mode, and then to switch to the
decision-directed (DD) mode. The proposed SG algorithm is obtained
by substituting the gradient terms (\ref{eq:proberror2}) and
(\ref{eq:proberror3}) in the expressions $\mathbf{
\bar{w}}_{k}(i+1)=\mathbf{ \bar{w}}_{k}(i)-\mu_{w}\frac{\partial
P_{e}}{\partial \mathbf{ \bar{w}}^{*}_{k}}$ and $\mathbf{
S}_{D}(i+1)=\mathbf{ S}_{D}(i)-\mu_{S_{D}}\frac{\partial
P_{e}}{\partial \mathbf{S}^{*}_{D}}$ \cite{haykin} subject to the
constraint of $\mathbf{ \bar{w}}^{H}_{k}(i)\mathbf{ S}_{D}^{H}(i)
\mathbf{ S}_{D}(i)\mathbf{ \bar{w}}_{k}(i)=1$. Based on
\cite{mber2}, we can see that, with respect to the product $\mathbf{
S}_{D}\mathbf{ \bar{w}}_{k}$, there are only global minimum
solutions, and all the solutions form a half hyperplane. In this
work, we pick one of the MBER solutions for $\mathbf{ S}_{D}\mathbf{
\bar{w}}_{k}$, which is with the unit length.
%
%
At each time instant, the weights of the two
quantities are updated in an alternating way by using the following
equations
\begin{equation}
\begin{split}
\mathbf {\bar{w}}_{k}(i+1)&=\mathbf{
\bar{w}}_{k}(i)+\mu_{w}\frac{\exp
\bigg(\frac{-|\Re[\bar{x}_{k}(i)]|^2}{2\rho^2} \bigg)\textrm{sign}
\{ b_{k}(i)\}}{2\sqrt{2\pi}\rho}
\\&\times \big( \mathbf{ S}^{ H}_{D}(i)\mathbf{ r}(i)-\Re[\bar{x}_{k}(i)]
\mathbf{ S}^{ H}_{D}(i)\mathbf{ S}_{ D}(i)\mathbf{
\bar{w}}_{k}(i)\big)\label{eq:proberror6}
\end{split}
\end{equation}
\begin{equation}
\begin{split}
\mathbf{ S}_{D}(i+1)&= \mathbf{ S}_{D}(i)+\mu_{S_{D}}
\frac{\exp \bigg(\frac{-|\Re[\bar{x}_{k}(i)]|^2}{2\rho^2}
\bigg)\textrm {sign}\{ b_{k}(i)\}}{2\sqrt{2\pi}\rho}
\\&\times \big(\mathbf{ r}(i)\mathbf{ \bar{w}}^{H}_{k}(i)-\mathbf{ S}_{D}(i)\mathbf{ \bar{w}}_{k}(i)
\mathbf{ \bar{w}}^{H}_{k}(i)\Re[\bar{x}_{k}(i)]
\big)\label{eq:proberror7}
\end{split}
\end{equation}
where $\mu_{w}$ and $\mu_{S_{D}}$ are the step-size values.
Expressions (\ref{eq:proberror6}) and (\ref{eq:proberror7}) need
initial values, $\mathbf {\bar{w}}_{k}(0)$ and $\mathbf
{S}_{D}(0)$, and we scale the reduced-rank filter by $\mathbf{
\bar{w}}_{k}=\frac{\mathbf{ \bar{w}}_{k}}{\sqrt{\mathbf{
\bar{w}}^{H}_{k}\mathbf{ S}_{D}^{H}\mathbf{ S}_{D}\mathbf{
\bar{w}}_{k}}}$ at each iteration.  The scaling has
an equivalent performance to using a constrained optimization with
Lagrange multipliers although it is computationally simpler.
The proposed adaptive
JIO-MBER algorithm is summarized in table \ref{tab:table1}.

\begin{table}[h]
\centering
 \caption{\normalsize Proposed adaptive JIO-MBER algorithms} {
\begin{tabular}{ll}
\hline
 $1$ & Initialize $\mathbf {\bar{w}}_{k}(0)$ and $\mathbf {S}_{D}(0)$.\\
 $2$ & Set step-size values $\mu_{w}$ and $\mu_{S_{D}}$\\
 $3$ &   for each time instant $i=0,1,\cdots$ do \\
 $4$ & ~~~~~~  Compute $\mathbf {\bar{w}}_{k}(i+1)$ and $\mathbf {S}_{D}(i+1)$ using (\ref{eq:proberror6}) and (\ref{eq:proberror7}). \\
 $5$ & ~~~~~~  Scale the $\mathbf {\bar{w}}_{k}$ using
 $\mathbf {\bar{w}}_{k}=\frac{\mathbf {\bar{w}}_{k}}{\sqrt{\mathbf {\bar{w}}^{H}_{k}\mathbf {S}^{H}_{D}\mathbf {S}_{D}\mathbf {\bar{w}}_{k}}}$. \\
 $6$ &
 Obtain
$\mathbf {\bar{w}}_{k}(i+1)$
and $\mathbf {S}_{D}(i+1)$ for the next
time instant. \\
\hline
\label{tab:table1}
\end{tabular}
}
\end{table}

 The joint optimization of $\mathbf{ \bar{w}}_{k}$ and
$\mathbf{ S}_{D}$ has been shown to converge to the global minimum
when the MSE is employed as the cost function \cite{delamaretvt10}.
The proposed scheme promotes an iterative exchange of information
between the transformation matrix and the reduced-rank filter, which
leads to improved convergence and tracking performance. However,
when the BER is used as the cost function, there are local minima
associated with the optimization.

\subsection{Computational Complexity of Algorithms}
We describe the computational complexity of the proposed JIO-MBER
adaptive algorithm in multiuser MIMO systems. In Table \ref{tab:table2},
 we
compute the number of additions and multiplications to compare the
complexity of the proposed JIO-MBER algorithm with the conventional  adaptive
reduced-rank algorithms, the adaptive least mean squares
(LMS) full-rank algorithm based on the MSE criterion \cite{haykin} and the SG
full-rank algorithm based on the BER criterion \cite{mber2}.
 Note that the
MWF-MBER algorithm corresponds to the use of the procedure in
\cite{MWF2} to construct $\mathbf{ S}_D(i)$ and
(\ref{eq:proberror6})  to compute $\mathbf{ \bar{w}}_{k}(i)$.
In particular, for a configuration with $M=32$ and $D=6$,
the number of multiplications for the MWF-MBER and the proposed
JIO-MBER algorithms  are $7836$ and $1225$, respectively. The number
of additions for them are $5517$ and $933$, respectively. Compared
to the MWF-MBER algorithm, the JIO-MBER algorithm reduces the
computational complexity significantly.

\begin{table}[h]
\centering%
\caption{\normalsize Computational complexity of Algorithms.} {
\begin{tabular}{ccc}
\hline \rule{0cm}{2.5ex}&  \multicolumn{2}{c}{Number of operations
per  symbol} \\ \cline{2-3}
Algorithm & {Multiplications} & {Additions} \\
\hline \vspace{0.25em}
\emph{\small \bf Full-Rank-LMS}  & {\small $2M+1$} & {\small $2M$}  \\
\emph{\small \bf Full-Rank-MBER}  &  {\small $4M+1$} & {\small $4M-1$}  \\
\emph{\small \bf MWF-LMS } \cite{MWF} &  {\small $DM^2-M^2$} &  {\small $DM^2-M^2$}  \\
                           &   {\small $+2DM+4D+1$}  &  {\small $+3D-2$} \\
\emph{\small \bf  EIG} \cite{eign2} &   {\small $O(M^3)$} &  {\small $O(M^3)$}  \\
\emph{\small \bf  JIO-LMS} \cite{delamarespl07} &   {\small $3DM+M$} &  {\small $2DM+M$}  \\
                          &   {\small $+3D+6$}  &   {\small $+4D-2$} \\
\emph{\small \bf MWF-MBER} \cite{mber3} &  {\small $(D+1)M^2$} &  {\small $(D-1)M^2$} \\
                       &   {\small $+(3D+1)M+3D$}  &  {\small $+(2D-1)M$}          \\
                       &      {\small $+M+10$}           &      {\small $+2D+M+1$}              \\
\emph{\small \bf JIO-MBER} & {\small $6MD+5D$}  & {\small $5MD+D$} \\
                       &        {\small $+M+11$}             &     {\small $-M-1$}            \\
\hline
\label{tab:table2}
\end{tabular}
}
\end{table}
\subsection{Automatic Rank Selection}
The performance of reduced-rank algorithms depends on the rank $D$,
which motivates automatic rank selection schemes to choose the best
rank at each time instant \cite{MWF2,delamaretvt10,jidf}. Unlike
prior methods for rank selection, we develop a rank adaptation
algorithm based on the probability of error, which is given by
\begin{equation}
P_{D}(i)=Q \bigg( \frac{\textrm{sign} \{
b_{k}(i)\}\Re[\bar{x}_{k}^{D}(i)]}{\rho }
\bigg)\label{eq:proberrorcostf}
\end{equation}
where the receiver is subject to $\mathbf{ \bar{w}}^{H}_{k} \mathbf{
S}_{D}^{H}\mathbf{ S}_{D}\mathbf{ \bar{w}}_{k}=1$. For each time
instant, we adapt a reduced-rank filter $\mathbf {
\tilde{\bar{w}}}_{k}(i)$ and a projection matrix $\mathbf {
\tilde{S}}_{D}(i)$ with the maximum allowed rank $D_{\rm max}$, which can be
expressed as
\begin{equation}
\mathbf{ \tilde{\bar{w}}}_{k}(i)=[\tilde {\bar{w}}_{1}(i), \ldots,
\tilde{\bar{w}}_{D_{\rm min}}(i), \ldots, \tilde{\bar{w}}_{D_{\rm
max}}(i)]^{T}
\end{equation}
\begin{equation}
\mathbf{ \tilde{S}}_{D}(i)=\left[ \begin{array}{ccccc} \tilde{s}_{1,1}(i)  &  \ldots & \tilde{s}_{1,D_{\rm min}}(i)
& \ldots &\tilde{s}_{1,D_{\rm max}}(i) \\
  \vdots &  \vdots & \vdots & \vdots &\vdots\\
\tilde{s}_{M,1}(i)  &  \ldots & \tilde{s}_{M,D_{\rm min}}(i) & \ldots &\tilde{s}_{M,D_{\rm max}}(i)\\
\end{array} \right]
\end{equation}
where $D_{\rm min}$ and $D_{\rm max}$ are the minimum and maximum
ranks allowed for the reduced-rank filter, respectively. For each
symbol, we test the value of rank $D$ within the range, namely,
$D_{\rm min}\leq D \leq D_{\rm max}$. For each tested rank, we
substitute the filter $\mathbf{
\tilde{\bar{w}}}^{'}_{k}(i)=[\tilde{\bar{w}}_{1}(i), \ldots,
\tilde{\bar{w}}_{D}(i)]^{T}$  and the matrix
\begin{equation}
\mathbf{ \tilde{S}}^{'}_{D}(i)=\left[ \begin{array}{ccc} \tilde{s}_{1,1}(i)  &  \ldots & \tilde{s}_{1,D}(i)  \\
  \vdots &  \vdots & \vdots \\
\tilde{s}_{M,1}(i)  &  \ldots & \tilde{s}_{M,D}(i) \\
\end{array} \right]
\end{equation}
into (\ref{eq:proberrorcostf}) to obtain the probability of error $P_{D}(i)$.
 The optimum rank can be selected as
\begin{equation}
D_{\rm opt}(i)=\arg \min_{D \in \{D_{min},\ldots, D_{max}\}}
P_{D}(i).
\end{equation}
The proposed MBER automatic rank selection requires the operation
with $D_{\rm max}$ to calculate
\begin{equation}
\begin{split}
\bar{x}_{k}^{D}(i)& =\mathbf {\bar{w}}^{H}_{k} (\sum_{d=1}^{D_{\rm
min}} \mathbf {s}^{H}_{d}\mathbf {r}(i) {\bf v}_d + \ldots +
{\mathbf s}^{H}_{D_{\rm opt}}\mathbf {r}(i) {\bf v}_{D_{\rm opt}} \\&\quad+  \sum_{d=D_{\rm opt}+1}^{D_{\rm max}}{\bf
s}^{H}_{d} \mathbf {r}(i) {\bf v}_{d}),
\label{sumx}
\end{split}
\end{equation}
where ${\bf v}_d$ is a zero vector with a one in the $d$th position and $\mathbf {s}_{d}=[\tilde{s}_{1,d}(i) , \ldots, \tilde{s}_{M,d}(i) ]^{T}$.
A simple search over the values of $\bar{x}_{k}^{D}(i)$ and the selection of
the terms corresponding to $D_{\rm opt}$ and $P_{D_{\rm opt}}(i)$
are performed.

\section{simulations}
\label{Section5:simulations}

In this section, we evaluate the  performance of the proposed
JIO-MBER reduced-rank algorithms and compare them with existing
full-rank and reduced-rank algorithms. Monte-carlo simulations are
conducted to verify the effectiveness of the JIO-MBER adaptive
reduced-rank SG algorithms.
The number of receive antennas at the BS is $M=32$. The channel coefficient
$h_{k,f}(i)$ is computed according to the Jakes model \cite{rappaport}.
We optimized the parameters of the JIO-MBER adaptive reduced-rank SG
algorithms with step sizes $\mu_{w}=0.01$ and $\mu_{S_{D}}=0.025$.
The step sizes for LMS adaptive full rank, SG adaptive MBER  full
rank and the conventional adaptive reduced-rank techniques are
$0.085$, $0.05$ and $0.035$, respectively. The initial full rank and
reduced-rank filters are all zero vectors. The initial projection
matrix is given by
$\mathbf{S}_{D}(0)=[\mathbf{I}_{D},\mathbf{0}_{D\times (M-D)}]^{T}$.
The algorithms process $250$ symbols in TR and $1500$ symbols in DD.
We set $\rho=2\sigma$.
%

\begin{figure}[!hhh]
\centering \scalebox{0.42}{\includegraphics{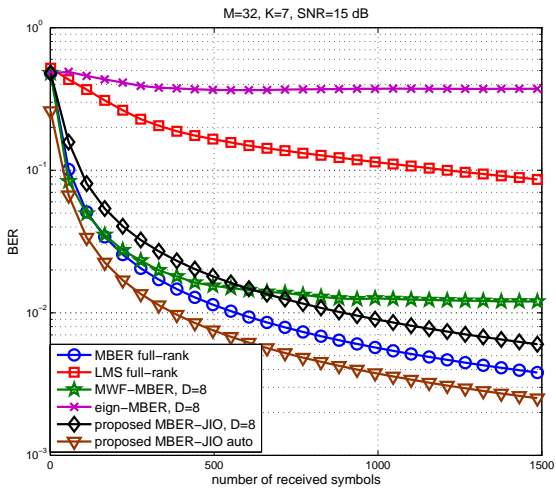}}
\vspace{-1.5em} \caption{BER performance versus the number of
received symbols for the JIO-MBER reduced-rank algorithms and the
conventional schemes. ($D_{min}=3$, $D_{max}=20$, $K=7$)  }
\label{fig:simulation1}
\end{figure}

\begin{figure}[!hhh]
\centering \scalebox{0.42}{\includegraphics{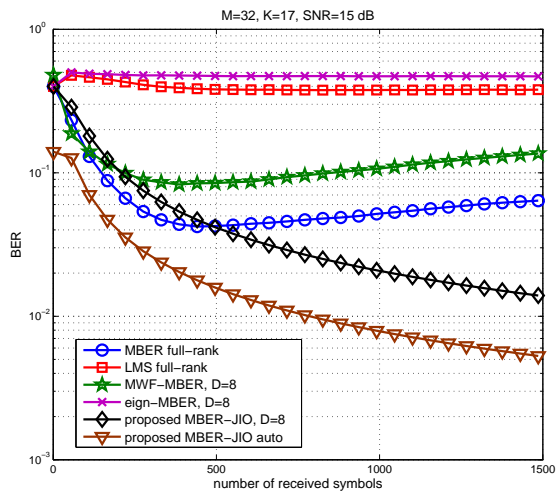}}
\vspace{-1.5em} \caption{BER performance versus the number of
received symbols for the JIO-MBER reduced-rank algorithms and the
conventional schemes. ($D_{min}=3$, $D_{max}=20$, $K=17$)  }
\label{fig:simulation2}
\end{figure}

Figs. \ref{fig:simulation1} and \ref{fig:simulation2} show the BER
performance of the desired user versus the number of received
symbols for the JIO-MBER adaptive SG algorithms and the conventional
schemes. We set the rank $D=8$ for the reduced-rank schemes,  and
the normalized  Doppler frequency is $f_{d}T_{s}=1\times10^{-5}$. We
use $15$ dB for the input signal to noise ratio (SNR). From Fig.
\ref{fig:simulation1} and \ref{fig:simulation2}, we can see that the
proposed JIO-MBER SG algorithm with the automatic rank selection
mechanism achieves the best performance.
Although the full-rank MBER SG algorithm has a better performance
compared to the proposed JIO-MBER SG algorithm with $D=8$ for a
system with a low load, the proposed JIO-MBER SG algorithm with
$D=8$ outperforms the full-rank MBER SG algorithm for a
highly-loaded system.
We also can see that the JIO-MBER reduced-rank algorithms converge
much faster than the conventional  reduced-rank algorithms, and the
MBER eigen-decomposition reduced-rank method with $D=8$ does not
work well in time-varying MIMO fading channels. For the group of
JIO-MBER adaptive algorithms, the auto-rank selection algorithms
outperform the fixed rank algorithms.

\begin{figure}[!hhh]
\centering \scalebox{0.45}{\includegraphics{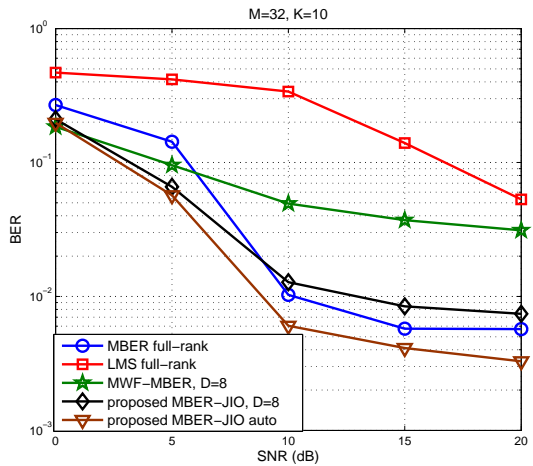}}
\vspace{-1.5em}\caption{ BER performance versus SNR for the JIO-MBER
reduced-rank algorithms and the conventional schemes. $1500$ symbols
are transmitted and $250$ symbols in TR. ($D_{min}=3$, $D_{max}=20$,
$K=10$) } \label{fig:simulation3}
\end{figure}

\begin{figure}[!hhh]
\centering \scalebox{0.45}{\includegraphics{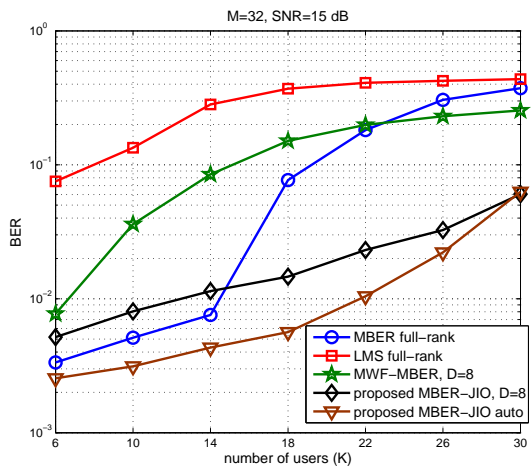}}
\vspace{-1.5em}\caption{ BER performance versus  number of users for
the JIO-MBER reduced-rank algorithms and the conventional schemes.
$1500$ symbols are transmitted and $250$ symbols in TR.
($D_{min}=3$, $D_{max}=20$, $SNR=15$ dB) } \label{fig:simulation4}
\end{figure}

Figs. \ref{fig:simulation3} and \ref{fig:simulation4} illustrate the
BER performance of the desired user versus SNR and number of users
$K$, where we set $f_{d}T_{s}=1\times10^{-5}$ and $D=8$.  We can see
that the best performance is achieved by the proposed JIO-MBER
algorithm with the automatic rank selection mechanism. The proposed
JIO-MBER algorithm with $D=8$ outperforms the MWF-MBER reduced-rank
algorithm. For the low-SNR region and the high-load case, the
proposed JIO-MBER algorithm with $D=8$ outperforms the full-rank
MBER SG algorithm.
%
%
In particular, the  JIO-MBER algorithm using the automatic rank
selection mechanism  can save up to over $5$dB and support up to six
more users in comparison with the full rank MBER SG algorithm, at
the BER level of $6\times 10 ^{-3}$.

\begin{figure}[!hhh]
\centering \scalebox{0.45}{\includegraphics{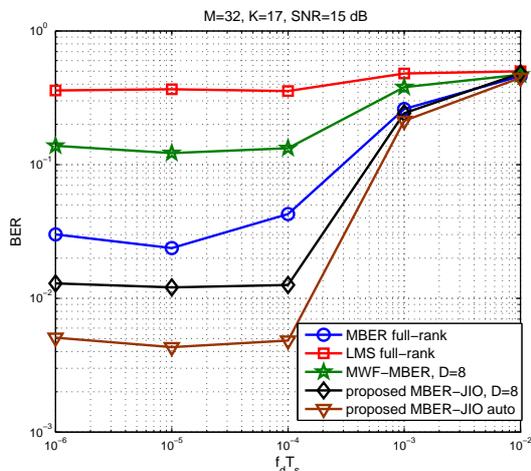}}
\vspace{-1.5em}\caption{ BER performance versus the (cycles/symbol)
for the JIO-MBER reduced-rank algorithms and the conventional
schemes. $1500$ symbols are transmitted and $250$ symbols in TR.
($D_{min}=3$, $D_{max}=20$, $SNR=15$ dB, $K=17$) }
\label{fig:simulation5}
\end{figure}

We show the BER performance of the analyzed algorithms as the fading
rate of the channels vary. In this experiment, we set the number of
users $K=17$ and $SNR=15 dB$. In Fig. \ref{fig:simulation5}, we can
see that, as the fading rate increases the performance gets worse,
and the proposed JIO-MBER algorithm with the automatic rank
selection mechanism achieves the best performance, followed by the
proposed JIO-MBER algorithm with $D=8$, the full-rank MBER SG
algorithm, the conventional MWF-MBER algorithm and the full-rank LMS
algorithm. It shows the ability of the proposed JIO-MBER algorithms
to deal with dynamic channels.

\section{Conclusions}
\label{Section6:conclusion}

In this paper, we have proposed a novel adaptive MBER reduced-rank
scheme based on joint iterative optimization of filters for
 multiuser MIMO systems. We have developed SG-based algorithms for
the adaptive estimation of the reduced-rank filter and the
projection matrix, and proposed an automatic rank selection scheme
using the BER as a criterion. The simulation results have shown that
the proposed JIO-MBER adaptive reduced-rank algorithms significantly
outperform the existing full-rank and reduced-rank algorithms at a
low cost.
\end{document}